\documentclass{aa}  

\usepackage{graphicx}
\usepackage{txfonts}
\begin{document}

   \title{The puzzling case of the radio-loud QSO 3C~186: a gravitational wave recoiling black hole in a young radio source?}


   \author{M. Chiaberge
          \inst{1,2}
          \and
          J.~C. Ely
          \inst{1}
          \and
          E.~T. Meyer
          \inst{3}
          \and
          M. Georganopoulos
          \inst{3,4}
          \and
          A. Marinucci
          \inst{5}
          \and
          S. Bianchi
          \inst{5}
          G.~R. Tremblay
          \inst{6}
          \and
          B. Hilbert
          \inst{1}
          \and
          J.~P. Kotyla
          \inst{1}
          \and
          A. Capetti
          \inst{7}
          \and
          S.~A. Baum
          \inst{8,9}
          \and
          F.~D. Macchetto
          \inst{1}
          \and
          G. Miley
          \inst{10}
          \and
          C.~P. O'Dea
          \inst{8,9}
          \and
          E.~S. Perlman
          \inst{11}
          \and 
          W.~B. Sparks
          \inst{1}
          \and
          C. Norman
          \inst{1,2}
          }

   \institute{Space Telescope Science Institute, 3700 San Martin Dr., Baltimore MD, 21210, USA\\
              \email{marcoc@stsci.edu}
         \and
             Johns Hopkins University, 3400 N. Charles Street, Baltimore, MD 21218, USA\\
         \and
         University of Maryland Baltimore County, 1000 Hilltop Circle, Baltimore, MD 21250, USA\\
         \and 
         NASA Goddard Space Flight Center, 8800 Greenbelt Road, Greenbelt, MD 20771, USA\\
         \and
         Dipartimento di Matematica e Fisica, Università degli Studi Roma Tre, via della Vasca Navale 84, 00146 Roma, Italy\\
         \and
         Department of Physics and Yale Center for Astronomy \& Astrophysics, Yale University, 217 Prospect Street, New Haven, CT 06511, USA\\
         \and
         INAF - Osservatorio Astrofisico di Torino, via Osservatorio 20, 10025, Pino Torinese, Italy\\
         \and
         University of Manitoba, Dept of Physics and Astronomy, Winnipeg, MB R3T 2N2, Canada\\
         \and 
         School of Physics \& Astronomy, Rochester Institute of Technology, 84 Lomb Memorial Dr., Rochester, NY 14623  USA\\
         \and
         Leiden Observatory, University of Leiden, P.O. Box 9513, 2300 RA Leiden, Netherlands\\
         \and
         Florida Institute of Technology, Physics \& Space Science Department, 150 West University Boulevard, Melbourne, FL 32901, USA            }

   \date{}

 
  \abstract
      {Radio-loud AGNs with powerful relativistic jets are thought to be associated with rapidly spinning black holes (BHs). BH spin-up may result from a number of processes, including accretion of matter onto the BH itself, and catastrophic events such as BH-BH mergers.}
   {We study the intriguing properties of the powerful (L$_{\rm bol}\sim 10^{47}$ erg  s$^{-1}$) radio-loud quasar 3C~186. This object shows peculiar features both in the images and in the spectra.}
   {We utilize near-IR {\it Hubble} Space Telescope (HST) images to study the properties of the host galaxy, and HST UV and Sloan Digital Sky Survey optical spectra to study the kinematics of the source. Chandra X-ray data are also used to better constrain the physical interpretation.}
   {HST imaging shows that the active nucleus is offset by 1.3 $\pm$ 0.1 arcsec (i.e. $\sim$ 11 kpc) with respect to the center of the host galaxy. Spectroscopic data show that the broad emission lines are offset by $-2140\pm390$ km/s with respect to the narrow lines. Velocity shifts are often seen in QSO spectra, in particular in high-ionization broad emission lines. The host galaxy of the quasar displays a distorted morphology with possible tidal features that are typical of the late stages of a galaxy merger.}
   {A number of scenarios can be envisaged to account for the observed features. While the presence of a peculiar outflow cannot be completely ruled out, all of the observed features are consistent with those expected if the QSO is associated with a gravitational wave (GW) recoiling BH. Future detailed studies of this object will allow us to confirm this type of scenario and will enable a better understanding of both the physics of BH-BH mergers and the phenomena associated with the emission of GW from astrophysical sources.}

   \keywords{Galaxies: active --
                quasars: individual: 3C~186 -- Galaxies: jets --
                Gravitational Waves
               }
   \titlerunning{A gravitational wave recoiling black hole in a young radio source?}

   \maketitle
%

\section{Introduction}

Radio-loud AGNs have been shown to be closely associated with galaxy major mergers \citep[][and references therein]{cliverev16}.
Mergers are expected to play an important role in the evolution of galaxies. These events may trigger star formation,
and may contribute to channel dust and gas towards the center of the
gravitational potential of the merged galaxy, where a supermassive black hole (BH) sits. This matter may ultimately  form
an accretion disk and turn-on an AGN (active galactic nucleus).
While this might not be the ultimate triggering mechanism for all AGNs, studying the properties of single objects at a great level of detail may help us to better understand the physical mechanisms at work in the vicinity of the central
supermassive black hole. 

When two galaxies that 
contain a supermassive black hole (SMBH) at their center merge,
the SMBHs are pulled towards the center of the gravitational potential of the merged galaxy 
by dynamical friction, and then
rapidly form a BH binary by losing angular momentum via gravitational slingshot interaction
with stars \citep{begelman80}. A few cases of SMBH binaries and dual AGN have in fact been observed 
\citep[e.g.][]{komossa03,bianchi08,deane14,comerford15}.
The third phase involves the emission of
gravitational waves, by which the bound BH pair may lose the remaining angular momentum, and eventually 
coalesce. How the two BHs reach the distance at which GW
emission becomes important is a process that is still poorly  understood, and it is possible that the 
binary may stall.
This is the so-called final parsec problem \citep{milosavljevicmerrit03}.
However, a gas-rich environment may significantly help to overcome this problem.  
Recent work using simulations also show that even in gas-poor environments 
SMBH binaries can merge under certain conditions, e.g. 
if they formed in major galaxy mergers where the final galaxy 
is non-spherical \citep[][and references therein]{khan11,preto11,khan12,bortolas16}.

When BHs merge, a number of phenomena are expected to happen. For example, the spin of the merged BH may be larger than
the initial spins of the two BHs involved in the merger. This strongly depends on the BH mass ratio and on the relative
orientation of the spins \citep[e.g.][for a recent review]{schnittman13}.
Recoiling  black holes  (BH) may also result  from BH-BH  mergers and  the
associated  anisotropic emission  of  gravitational  waves \citep[GW,][]{peres62,beckenstein73}. 
The resultant  merged BH may receive a kick
and  be displaced  or even  ejected from  the host  galaxy \citep{merritt04,madauquataert04,komossa12}, 
a process  that has
been  extensively studied  with simulations  \citep{campanelli07,blecha11,blecha16}. 
Typically, for non-spinning BHs, the expected velocity is
of the order of a few hundreds of km s$^{-1}$, or less. 
Recent work based on numerical relativity simulations have shown that
 superkicks of up to $\sim 5000$ km s$^{-1}$ are possible, but are
expected to be rare \citep[e.g.][]{campanelli07,bruegmann08}.

Emission of gravitational waves from merging SMBH may be detected in the future 
with space-based detectors such as LISA. For the most massive BHs (M$_{\rm BH} > 10^7$ M$_{\odot}$)
the frequency of the emitted GWs is low enough to allow detection with
pulsar-timing array experiments \citep[e.g.][and references therein]{sesanavecchio10,moore15}.
Finding  evidence for 
BHs that were  ejected from their post-merger single host galaxy center is extremely important to both
test the theory of GW kicks  and even more fundamentally to prove that
supermassive BH mergers  do occur.

If the ejected merged BH  is active, we
expect to observe an offset nucleus and velocity shifts between narrow
and broad lines  \citep{loeb07,volonterimadau08}. Such an offset is expected because 
the broad-line emitting region is dragged out with the kicked BH, while the narrow-line region is not.
However, because spectral lines of QSOs often show relatively large shifts \citep[$\sim$ a few hundred km/s,][]{shen16}, it
is extremely hard to properly model those spectra and identify true GW recoiling BH  candidates.
In fact, a few  candidates have been
reported so far in the literature, but equally plausible  alternative interpretations exist for
these observations. 
In general, no conclusively proved case of a GW recoiling black hole has been found so far, 
since it is difficult to disprove alternative explanations.

One  of the most convincing cases  reported so far
is  the  merging galaxy  CID-42  \citep{civano10,civano12,novak15}. 
This object shows two galaxy nuclei, one of which contains a point source associated with a
broad-lined AGN. The broad H$\beta$ emission line in this AGN is significantly offset ($\sim 1300$ km s$^{-1}$) with
respect to the narrow line system.
However,  alternative explanations such as  a dual-AGN scenario
\citep{comerford09} are still viable. Other interesting candidates
that show offset nuclei include NGC  3718 \citep{markakis15}, the 
quasar SDSS 0956+5128  \citep{steinhardt12}  and   SDSS  1133   \citep{koss14}. 

Low-luminosity radio-loud AGNs (RLAGN) that only show small
spatial offsets between the active nucleus and the isophotal center of the host galaxy 
\citep[$<$  10 pc,][]{batcheldor10,lena14} have also been found.
In addition, a few  objects that show velocity offsets, but for which 
evidence for spatial offsets is yet to be found, are also present
\citep{eracleous12,kim16}. But the  best GW recoiling BH candidates are
those that show both of these properties \citep{blecha16}.

Here  we present  evidence for  both spatial  and velocity  offsets in
3C~186, a  young \citep[$\sim 10^5$ yr,][]{murgia99} RLAGN  that belongs to
the   compact-steep  spectrum   class  \citep{fanti85,odea98}.  
3C~186  is  located  in  a  well-studied  cluster  of  galaxies
\citep{siemiginowska05,siemiginowska10}.  Its redshift, as measured by \citet{hewittwild10} using the
Sloan  Digital Sky  Survey  Data  Release 6  (SDSS  DR6), is  z=1.0686
+/-0.0004. We show that, although alternative interpretations
cannot be completely excluded, a scenario involving a GW recoiling BH is viable.

The structure of this paper is as follows. In Sect.~\ref{observations} we describe the datasets; in Sect.~\ref{analysis}
we outline the steps of the data analysis and we show results; in Sect.~\ref{discussion} we discuss 
possible interpretations for our findings. Finally in Sect.~\ref{conclusions} we draw conclusions
and we outline future work.

The AB magnitude system and the following cosmological parameters are 
used throughout the paper: H$_0$=69.6 km/s/Mpc, $\Omega_M$=0.286, $\Omega_\lambda$=0.714.

\section{Observations}
\label{observations}
\subsection{{\it Hubble} Space Telescope imaging}
We obtained  Hubble Space  Telescope (HST) images  of 3C~186  using the
Wide  Field Camera  3 (WFC3)  as  part of  our Cycle  20 HST  SNAPSHOT
program GO13023. Images in the rest-frame optical and UV taken with 
the IR and UVIS channels, respectively, are described
in detail in \cite{hilbert16}  for the full sample. In this paper
we only use the WFC3-IR F140W image. This filter is centered at 1392nm
and has  a width  of 384nm.  Two dithered images  were taken  and then
combined using Astrodrizzle (Fruchter et al. 2012). The total exposure
time is 498.5 seconds. The UVIS F606W image does not add any 
significant information 
to the analysis presented in this paper. In fact, in the region of interest it only shows the quasar point source 
and a blob of uncertain origin, located $\sim 2$'' East-North-East of the QSO (Fig.~\ref{fig1}, top-left panel).

\subsection{Spectroscopy}
The  UV and  optical spectroscopic  data are  from HST  and the  Sloan
Digital Sky  Survey (SDSS), respectively.  The HST spectrum  was taken
with the Faint  Object Spectrograph (FOS) as part  of program GO-2578.
The data were taken in 1991  using the G270H and G400H gratings, which
span the  wavelength range  from 2221 to  4822\AA. The  total exposure
time  is   1080  seconds  and   846  seconds  for  G270H   and  G400H,
respectively. The  SDSS observations were  taken in 2000,  using plate
433 and fiber  181. The datasets were used as  delivered from the MAST
(Mikulski Archive  for Space  Telescopes) and  from the  SDSS archive,
with no post-processing applied.

\subsection{X-ray {\it Chandra} data}

3C 186 was observed five times  (Siemiginowska et al. 2005, 2010) with
the Chandra X-ray Observatory with  the ACIS-S detector. We merged the
last four observations, which were all performed in December 2007. The
resulting total  exposure time  is 197ks. Data  were reduced  with the
Chandra  Interactive  Analysis  of  Observations 4.7  and  the  latest
Chandra Calibration Database (CALDB), adopting standard procedures.

\section{Data analysis and results}
\label{analysis}
\subsection{HST imaging}
\label{hstimage}

\begin{figure*}
\centering
\includegraphics[width=18cm]{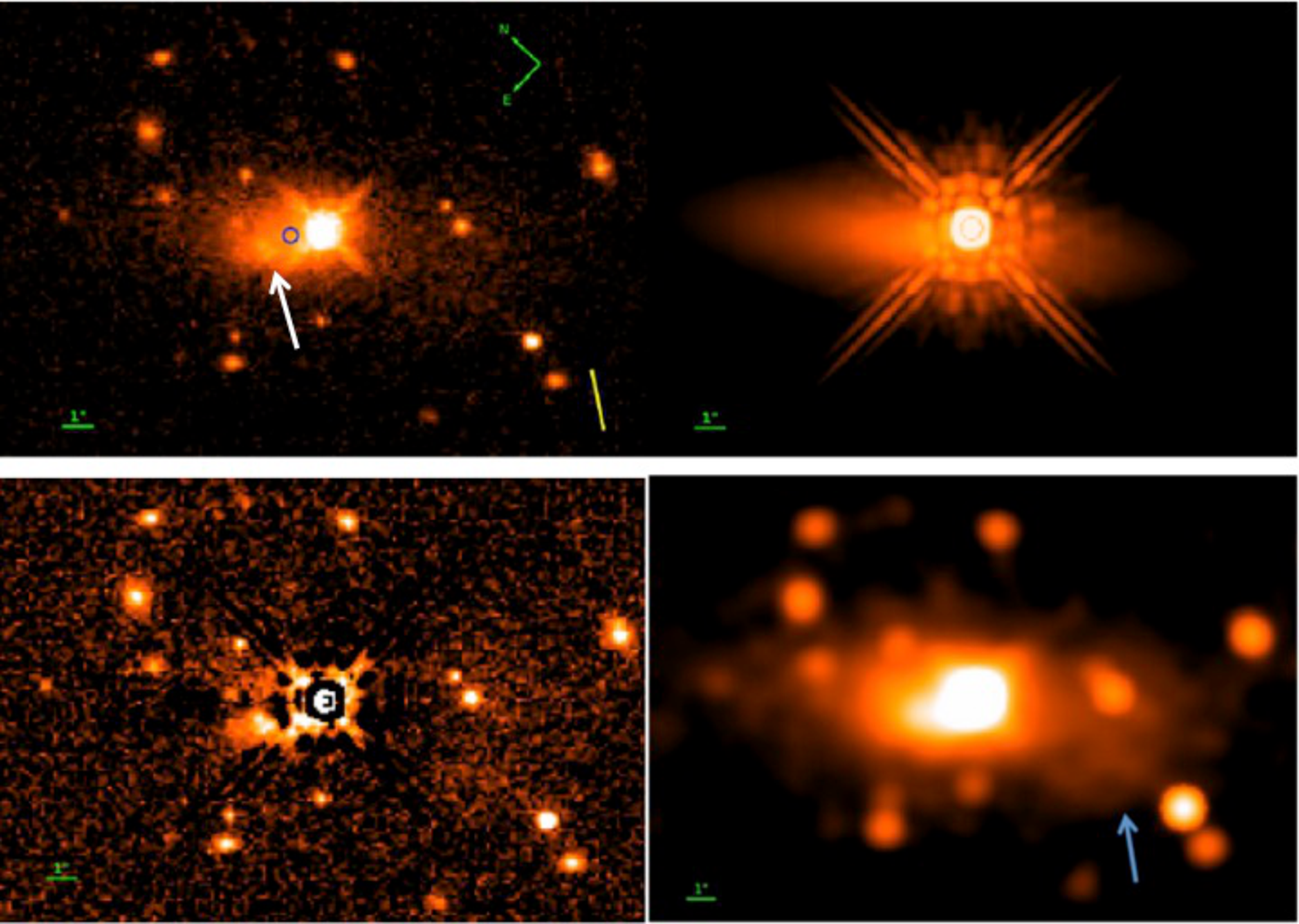}
\caption{HST image of 3C~186 (top-left). The host galaxy center is indicated with a blue circle. The orientation of the radio jet is shown as a yellow line. The white arrow indicates the location of the so-called blob of unknown origin, $\sim 2$ arcsec East-North-East of the quasar point source. Top-right: model of the source, which includes a PSF and a S\'ersic model. Bottom-left: residuals after model subtraction. Bottom right: smoothed (4-pixel kernel) version of the HST image showing the presence of low S/N shells or tidal tails in the host galaxy (indicated by the blue arrow). }
\label{fig1}
\end{figure*}

\begin{table}
\caption{2-D modeling best fit parameters}             
\label{tab1}      
\centering                          
\begin{tabular}{c c c c c c}        
\hline\hline                 
~ & mag$_{\rm F140W}$ & r$_{\rm eff}$ & n & e & P.A. \\
~ & (1) & (2) & (3) & (4) & (5) \\    
\hline                        
Sérsic & 18.86	& 6.47”	& 2.57	& 0.25	& 37.28 \\
err. &	.06	& 0.61”	& 0.20	& 0.01	& 0.41 \\
\hline
PSF &	17.39	& --	& --	& --	& -- \\
err. &	0.01	& --	& --	& --	& -- \\
\hline                                   
\end{tabular}
\tablefoot{The reduced $\chi^2$ is 1.274. The reported parameters are the magnitude (1) the effective radius in arcseconds (2), the S\'ersic index (3), the ellipticity (4) and the position angle relative to the North (5). Errors are derived from
{\it Galfit}.}
\end{table}

We  fit  the  HST  image  (Fig.  1,  top-left  panel)  using  the  2-D
galaxy-fitting  algorithm {\it Galfit}  (Peng et  al. 2010).  Two components
are used  in the  fit: i)  a point-spread function  (PSF) to  fit the
quasar, and  ii) a galaxy  profile with  a Sérsic function  (Sersic et
al.  1963) for  the host  galaxy.  We  use an  undistorted PSF  model
derived with  {\it Tinytim} (Krist et  al. 2011) that was calculated  using different
power-law spectra  with slopes ranging from  0.3 to -1 (F$_\nu  \propto \nu^{-\alpha}$) and
performing different  focus corrections (from f=-0.24  to f=0.91). The
observed residuals are  only weakly dependent on  these parameters.
Using  both the $\chi^2$  and visual inspection of  the residuals, we determine
that  the  best  results  are  obtained using  $\alpha$=0  and  f=0.91.   The
undistorted PSF  model image is  oversampled by  a factor of  1.3 with
respect to the  original pixel size, therefore we  resampled the image
on the same  pixel scale using Astrodrizzle. The {\it Tinytim}  model is not
optimal, especially for  the core of the PSF, but  using a PSF derived
from observations of stars in  the WFC3 PSF library database (Anderson
et al. 2015) does  not produce better results in terms  of both $\chi^2$ and
residuals.

We mask out additional sources in the  field of view, in a region of
about 10$\arcsec$ radius centered on the quasar. Most of these are 
likely small cluster galaxies at the redshift of the target.
Masking out those objects has a significant effect
on the  output magnitude  of the  host galaxy only,  while the  point
source  flux  and  position  of both  components  are  unchanged.  The
best-fit model  parameters are reported  in Table 1. The 2-D model and  the residuals
after model  subtraction are shown  in Fig. 1, top-right and bottom-left  panels, respectively. 
One residual  blob   is  visible   $\sim 2\arcsec$  East-Northeast  of   the  quasar
center.  This feature was also masked out during the fitting process. 
Its origin is not established  but, owing to its very blue color, it 
may  possibly be a region of intense star formation, as discussed in \citet{hilbert16}.

To  obtain reliable estimates of the errors  on the important
parameters,  we  fixed some  of  the  parameters  to values  that  are
slightly different from  the best-fit value and we  checked the effect
on  both the  $\chi^2$  and  the residuals.  We  conclude  that the  largest
uncertainty is in the Sérsic index. If varied between n=1.9 and
n=3.7,  no effect  on  the $\chi^2$ is seen and  very limited changes  in  the
residuals are observed. 
The results of the analysis show that the quasar PSF
is not located  at the center of the host  galaxy. The offset measured
from the  best fit is 1.32 $\pm$ 0.05 arcsec, which corresponds to  a projected
distance of ~11  kpc, at the redshift of the  source, assuming a scale
of 8.244 kpc/arcsec. We also find that fixing the center of the host 
galaxy to the center of the PSF results in a statistically significant worse fit.
In fact, in this case the reduced $\chi^2$ increases to 1.292 and the $\chi^2$ 
difference test returns a probability P $<0.005$ than the two fits are the same.
Therefore, we conclude that the offset is real.

We note that the effective radius we derive (6.47”, corresponding to 53 kpc at the redshift of the target) is close to the average radius of other well studied BCGs in the same redshift range \citep[i.e. r$_{\rm eff} = 57.3 \pm 15.7$ kpc,][]{stott11}.

We also note that the host galaxy shows the  presence of low surface brightness features that extend 
to $\sim 6$'' south-east of the center of the QSO (Fig.~\ref{fig1}, bottom-right panel). These are  possibly  shells or  tidal  tails  that  are
typically associated with remnants of galaxy major  mergers  (Fig.~\ref{fig1},  bottom-right
panel). Those regions are irrelevant with respect to determining the host galaxy center,
because of their extremely low surface brightness. We tried to include a second large-scale component
to model this area of the host, but {\it Galfit} does not find any meaningful solution. Furthermore, even allowing for the presence of such an additional component, the derived center
of the host galaxy is still located at the position derived with the single S\'ersic + PSF model discussed above.

\subsubsection{Black hole mass estimate}
\label{bhmass}
Allowing for some  level of uncertainty (typically a  factor of $\sim$3),
we may infer the  mass of the BH by using  specific properties of the
host galaxy  (i.e. stellar  central velocity  dispersion, bulge  luminosity,
stellar mass)  as indicators. The magnitude of the host galaxy of  3C~186 as derived from our 2-D fit,
is m$_{\rm F140W}$=18.86 $\pm$ 0.06. Using the  WFC3 Exposure
Time Calculator  tool, we determine that this  corresponds to a near  IR K-corrected
K-band magnitude  K=17.1 (in the  Vega system), assuming  the spectral
energy distribution of  an elliptical galaxy. Using  the relation that
links the K-band magnitude to the BH  mass \citep{marconihunt03}, we obtain a BH mass of $3
\times 10^9$ solar  masses. This is the expected  mass of the 
SMBH associated with the galaxy we detect in the image.

In Sect.~\ref{nohost}  we also estimate the BH mass using the information derived from the spectra, 
and we will show that the two values are consistent with each other, within the errors. Furthermore, 
we use that information to set a tight constraint on the presence of any additional host galaxy around the QSO.
This enables us to determine that the host galaxy of the QSO is in fact the one we see in the HST image, which
is a very important piece of information to provide a consistent physical interpretation of the data.

\subsection{Spectroscopy}

\subsubsection{Spectral modeling}
\label{spectralmodeling}

\begin{figure*}
\includegraphics[width=18cm]{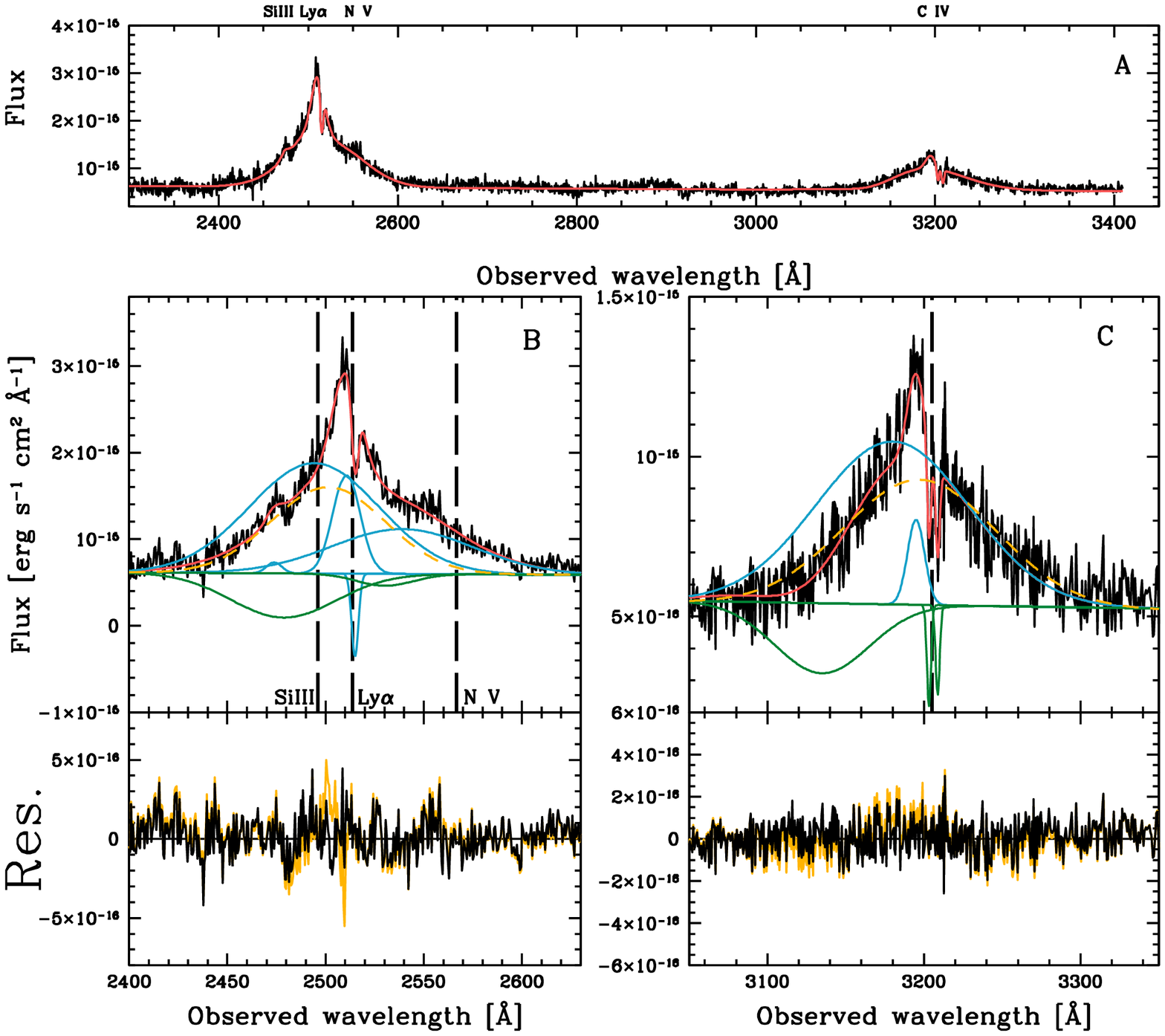}
\caption{HST/FOS UV spectrum. Wavelengths are in the observer’s frame. In the top panel (A) we show the full spectrum and the best-fit model (red line). Relevant lines are labeled on top of the panel, at the corresponding observed wavelength. The lower panels show zoomed-in regions for each of the lines discussed in the text. Panels B and C show the regions of the Ly$\alpha$ complex and  C~IV, respectively. The best fit is the red line. Each component of the model is shown separately, added to the continuum power law, for clarity. The emission components are shown in blue and the absorption components are shown in green. Broad components of the best fit derived without including broad absorption are shown as yellow dashed lines. The model residuals are also shown at the bottom of panels B and C. The yellow lines refer to the model without broad absorption. The thick dashed vertical lines correspond to the wavelength of each line at the systemic redshift measured from the narrow lines (see Fig.~\ref{fig3}, panel D).}
\label{fig2}
\end{figure*}

\begin{figure*}
\includegraphics[width=18cm]{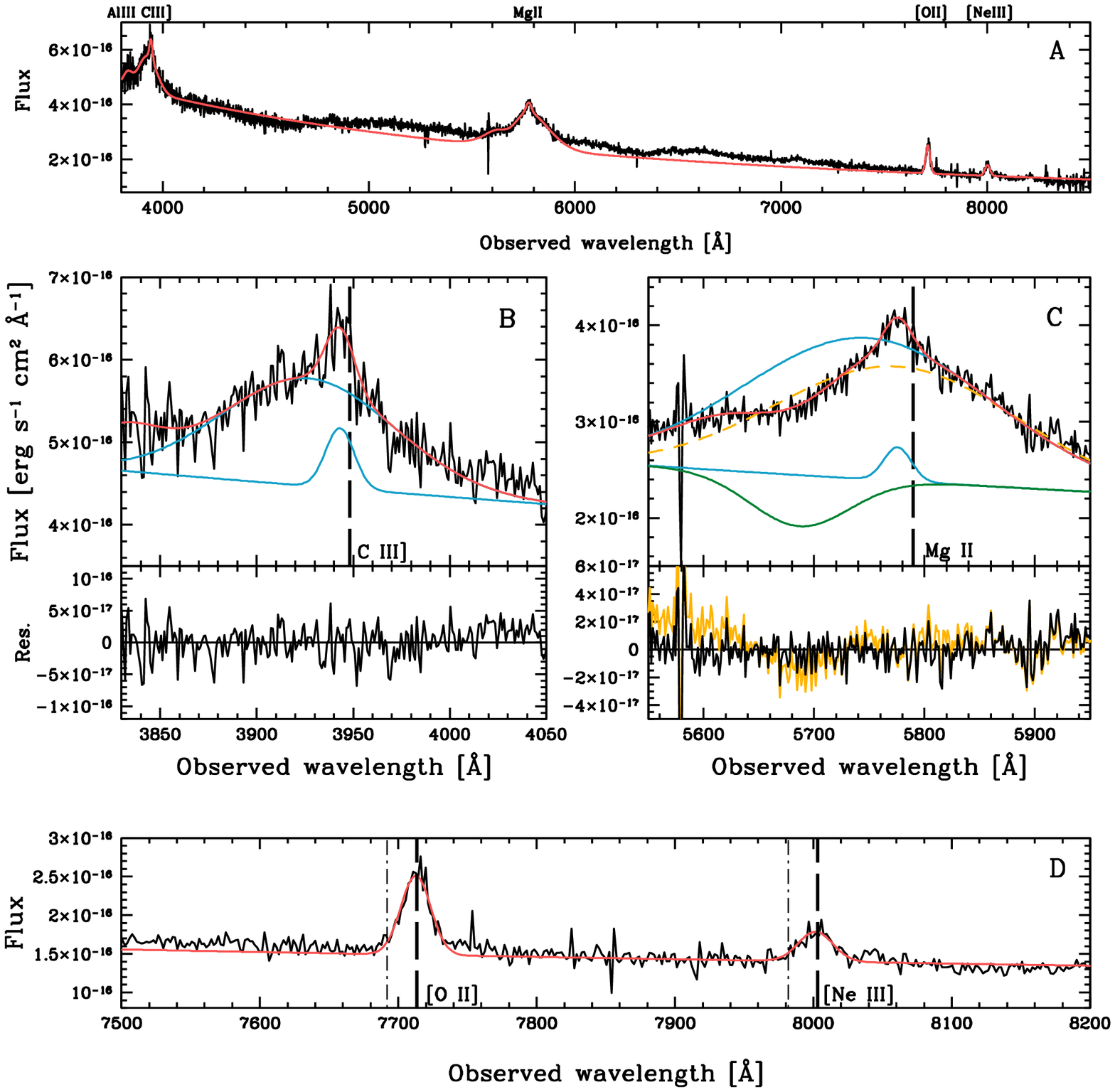}
\caption{Same as in Fig. 2 but for the SDSS optical spectrum. Panel B and C show the regions of C~III] and Mg~II, respectively.  Note that the continuum in the region of the Mg~II line (between $\sim$5000\AA and 7000\AA) is not reproduced by the fit because of the presence of Fe II features. The best fit is the red line. Each component of the model is also shown, as in Fig~\ref{fig2}. The emission components are shown in blue and the absorption components are shown in green. Broad components of the best fit derived without including broad absorption are shown as yellow dashed lines. The model residuals are also shown at the bottom of panels B and C. The yellow lines refer to the model without broad absorption. The thick dashed vertical lines correspond to the wavelength of each line at the systemic redshift measured from the narrow lines (see panel D).
Panel D shows the spectral region of the two isolated [O~II]3727 and [Ne~III]3869 narrow lines.  The dot-dashed lines in panel D indicate the wavelengths of these lines corresponding to the redshift of the source estimated by Kuraszkiewicz et al. (2002, see Sect.~\ref{kuraszkiewicz} for more details). }
\label{fig3}
\end{figure*}

Spectral fitting  is performed using  the {\it Specfit} tool in  IRAF.  The
spectrum is fit  with a global power-law and a  collection of Gaussian
profiles  to  each  line  of   interest.   The  parameters  are  then
successively  freed and  optimized through  a maximum of 100  iterations using a combination
of the {\it Simplex} and {\it Marquardt} minimization algorithms.  
The optimal
parameters  for  each  line  are  determined  until  convergence  is
achieved. 
The most  prominent features in  the HST FOS  UV spectrum are  Ly$\alpha$ and
C~IV1549  (Fig.~\ref{fig2}, panel A).  The  optical SDSS spectrum  shows C~III]1909,  Mg~II2798,
[O~II]3727 and [Ne~III]3869 (Fig.~\ref{fig3}, panel A).

The procedure used to derive the best-fit parameters is as follows.
We first fit each line complex separately, 
focusing on the spectral region dominated by each line, to limit the contamination from additional features. 
This is particularly important for Mg~II, to isolate such a line from 
the possible contamination from Fe~II features. 
At this first step we use the parameters for the continuum power law 
derived from a first-guess global fit.
The best-fit values found for each single line complex is then
used in the global fit as first guesses. The errors are estimated from the final global fit.

We checked that the spectral region between $\sim 5600\AA$ and $5700\AA$ (corresponding to a rest frame  wavelength range of $\sim 2710-2750\AA$) is not significantly contaminated by Fe~II emission. We followed the prescriptions of
\citet{vestergaardwilkes01}, i.e. we compared the continuum-subtracted emission of the Fe~II features in the pure iron spectral region between 2500 and 2600$\AA$ with the flux level measured in the above range of wavelengths. We derived that the flux immediately blue-ward of the peak of the Mg~II line is significantly higher than that expected from Fe~II features (by a factor of at least $\sim 2$). A larger contribution from Fe~II features is expected red-ward of the Mg~II line (around $\lambda_{obs} \sim 6100\AA $) and the observed features are consistent with the expectations in that spectral range.

Broad (FWHM  $>$ 3000 km/s) and  narrow (FWHM $<$ 3000  km/s) emission
and  absorption components are  used to  fit the  spectra.  The
[O~II] and [Ne~III] forbidden narrow lines are each fit with a
single  component.  Ly$\alpha$,  C~IV, C~III],  and Mg~II  are each  fit
using  broad  and  narrow emission  components.   Narrow  absorption
components  are  also  required  for both  Ly$\alpha$  and  the  C IV
doublet. 
For the Ly$\alpha$ complex, the presence of the Si~III 1206 line is also apparent, at an observed wavelength of $\sim 2475$\AA ~~(Fig.~\ref{fig2}, panel A). 
In the spectral model of the SDSS data we also include the Al~III 1857 line to
better reproduce the spectrum blue-ward of the C~III] line. This is purely done for cosmetic reasons, since the
extremely low S/N ratio at the blue edge of the SDSS spectrum does not allow a clear identification of such
a feature.

In addition,  for the permitted Ly$\alpha$, N~V, C~IV, and
Mg~II rest-frame UV lines, we include a broad absorption component in the spectral model.
Such a feature is possibly interpreted as being due to
a blue-shifted outflow. 
Fast, broad absorption features
have been recently observed in the UV spectra of a number of AGNs, most notably in NGC~5548 \citep{kaastra14} and 
NGC~985 \citep{ebrero16}.
In the following, we show that
while in principle these lines can be fitted without 
broad absorption, including such a component has the effect both of
improving the fit with a high statistical significance, and of providing
a physically consistent picture of all of the observed emission lines.

The use of a model that includes broad absorption is motivated by
the fact that  the profile of  the broad emission lines  appears strongly asymmetric,
especially for some of the detected lines. In particular, for both Mg~II and C~IV, the blue
side of the line is clearly concave. The same seems to hold for Ly$\alpha$,
but the presence of N~V and Si~III on the red and blue side of that line, respectively, 
makes the concave shape less obvious. 

While the depression observed blue-ward of the line peak might be a signature of
an intrinsic  asymmetry of the lines,  our choice to fit  the spectrum
with Gaussian components and  to include blue-shifted broad absorption
lines  is motivated  by the  following reasons:  i) the
asymmetry  is  particularly  strong  for all  the  resonant  lines, while there is  no evidence for any
asymmetries  in the  C~III] semi-forbidden  line, for  which we  do not
expect broad  absorption to be  observed; ii) by  utilizing Gaussian
lines and broad  absorption, we can fit all lines  with a consistent
symmetric profile. Instead,  if we were to  use asymmetric profiles,
each line  would have  a unique shape  that  would be difficult to interpret.  

Relevant line  parameters  derived  from the  best-fit
models are displayed in Table 2. In Figs.~\ref{fig2} and \ref{fig3}, panels B and C, we
show the best fit spectral model for each line complex (red line). Each of the Gaussian components of the model
are shown separately, added to the continuum, in blue and green for emission and absorption, respectively.

To assess the impact of our spectral model assumptions on
the results, we also fit the spectra without using a broad absorption component, 
and we compare the results by performing a $\chi^2$ difference test.
We simply run {\it Specfit} for each of the resonant lines separately, removing the broad absorption 
component from the fit and freeing all other parameters. The broad emission component derived with this spectral model is 
plotted in Figs.~\ref{fig2} and \ref{fig3} as a yellow dashed line. Then we compare the
value of $\chi^2$ with that obtained using the best fit (with broad absorption) for the same range of wavelengths.
For all lines the fit is significantly better when the broad absorption 
component is used. The significance level, given by the probability that the inclusion of the extra component
does not improve the fit, is P$<<$0.001  for both 
Ly$\alpha$ and Mg II, while for C IV
the significance is P$<$0.01.
In Figs.~\ref{fig2} and \ref{fig3}, we show a comparison of the residuals of the best fits obtained with and without
broad absorption (yellow lines in the residuals box of panels B and C of Fig.~\ref{fig2}, 
and in panel C of Fig.~\ref{fig3}). 
The improvement when broad absorption is used is obvious.
When using a model with no broad absorption, the worst fit is obtained in the case of Mg II, where the concavity of the
blue side of the line is particularly prominent. In that case, we also try using multiple
Gaussian emission components to achieve a better fit, but this type of model does not allow the the fit to converge.

\subsubsection{Spectral modeling results: evidence for velocity offsets}
\label{spectraresults}
The two isolated narrow  lines in the SDSS
spectrum ([O~II] and  [Ne~III], see Fig.~\ref{fig3}, panel D) are the best  features to derive
the value of the systemic redshift of the host galaxy, since these lines are produced in the narrow line region (NLR)  
on $\sim$kpc scales, far from the BH. 
The redshifts of  these lines are consistent  with each
other within  1$\sigma$. By averaging  the two redshifts we  derive z$_h$=1.0685
$\pm$ 0.0004. This is consistent with the literature value of z=1.0686 reported by NED \citep{hewittwild10}.

Strikingly, the FOS spectrum shows the presence of a narrow absorption line for Ly$\alpha$,
as well as the  C~IV 1548,1551\AA ~~absorption  doublet . The  redshift of
these three  narrow absorption lines  is consistent with  the systemic
redshift of the host z$_h$ derived from [O~II] and [Ne~III], within 1$\sigma$ and 2$\sigma$  for 
C~IV and  Ly$\alpha$, respectively.

All of the observed broad lines show a substantial offset (blue-shift)
with respect to the narrow line system. 
In Figs.~\ref{fig2} (panel B and C) and \ref{fig3} (panel C) we indicate the wavelengths corresponding to
the systemic redshift z$_h$ for each major emission line with thick black vertical dashed lines. This
shows the velocity offsets of the broad emission lines very clearly.
The offsets of all broad emission components of the major emission lines 
are consistent  with each  other within $\sim1\sigma$ (see Tab.~\ref{table2}). 

The measurement with the largest error is obtained for N~V, which 
is a very broad and relatively faint high ionization line that is heavily
blended in the Ly$\alpha$ complex. 
The results for Al~III are reported in Table~\ref{table2} only for the sake of 
completeness. Even if there is evidence for a significant offset, we believe that the derived 
value is not reliable for this line because of the
extremely low S/N ratio at the red end of the SDSS spectrum.

\begin{figure}
\includegraphics[width=9cm]{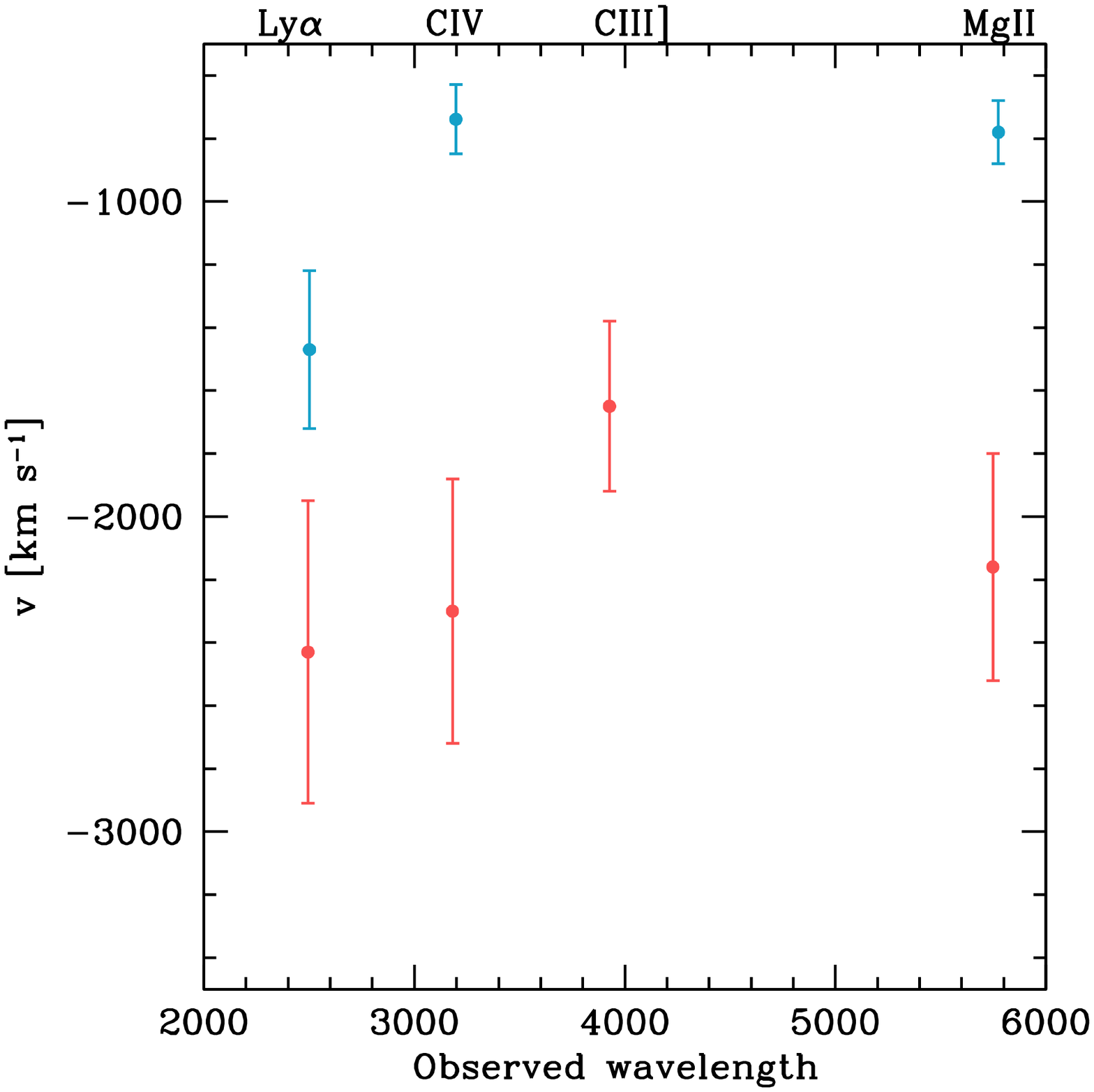}
\caption{Velocity offsets of the broad emission lines as measured with respect to the systemic redshift z$_h$ plotted
against the observed wavelength of each line. The light blue points and the red points are for the spectral models without and with broad absorption, respectively. Note that there is only one point representing the C~III] line because for semi-forbidden lines broad absorption is not expected.}
\label{fig4}
\end{figure}

We use the four strongest broad lines (i.e. Ly$\alpha$, C~IV, C~III] and Mg~II) to derive
the average\footnote{If we use all of the broad lines identified in the spectra, including the fainter 
Si~III, N~V and Al~III lines, the average velocity offset is v$=2190\pm 550$km s$^{-1}$.} velocity 
offset  v $= -2140 \pm 390$ km s$^{-1}$.
In Fig.~\ref{fig4} we plot the velocity offset against the central wavelength for each of the major broad lines.
The data points derived using a broad absorption component in the fit for the resonant lines 
are shown in red. In blue we show the velocity offsets derived without assuming the
presence of broad absorption.
We note that even without the inclusion
of the broad absorption component, and allowing for a less accurate fit, the broad emission lines are still
significantly offset with respect to the systemic wavelength, although the velocity offsets are smaller ($\sim$1000 km/s). However, the CIII] line is still significantly above that value, since no broad absorption is adopted in our analysis for that  non-resonant line.
Furthermore, we wish to point out that the resulting velocity offsets for each of the lines
are not consistent with 
each other in the case in which no broad absorption is included.
Therefore, we conclude that a model with broad absorption components
both produces a better representation of the data, and provides a physically consistent picture of the source.

To establish that the assumption of the Gaussian shape for all lines
is not artificially generating the line shifts, we  perform  a
measurement of the flux-weighted centroid  of the broad component of
the C~III] line. This is the only broad line included in the available spectra 
for which we do not expect broad absorption to 
significantly affect its shape.
We masked out both the  emission of the narrow component and the
region  blue-ward  of  C~III],  where Al~III  might  contaminate  the
continuum. Without  assuming any specific profile,  the broad line
is  centered  at 3920$\pm$15\AA,  corresponding  to  z=1.054, and  still
significantly offset  (v$\sim -2140$ km/s) with respect  to the systemic
redshift z$_h$.

We also note that each emission line complex is best fitted with the inclusion of a narrow component that is slightly blue-shifted
with respect to the systemic redshift. In Table~\ref{table2}, we report each of these lines with a question mark, 
since their
origin is not well determined. These components could be possibly be explained as being due to 
outflows of moderate velocity ($\sim$100 km s$^{-1}$) in the narrow line region. 

In Appendix ~\ref{palomar} we also include first results from a subset of data obtained at the Palomar Observatory 200'' telescope with TripleSpec.
  The only spectral region free of significant atmosphere absorption that includes broad emission lines shows that the He I line
  is best fitted using a broad absorption absorption component with the same properties as for the UV resonant lines. A more in-depth analysis of the
  full dataset will be presented in a forthcoming paper.

In summary, our analysis of the spectra show that 
we can identify two main systems at two different redshifts: the host galaxy (and the NLR) at z$\sim 1.068$, and the broad line region (BLR) at z$\sim 1.054$.

\begin{table*}
\caption{Emission lines best-fit model}             
\label{table2}      
\centering          
\begin{tabular}{l c c c c c c c c }     
\hline\hline       
Line component	      & Observed wavelength & Err.            &  Redshift &   Err. &  Velocity offset  &   Err. & FWHM &	Err \\
   ~                  & $\lambda$ [\AA])    &  ~ & $z$ & ~ & [km s$^{-1}$] & ~ & [km s$^{-1}$] \\
\hline  
\hline
[O~II]	                     &  7712.7	& 0.9  &   1.0686       &0.0002 & --	& --	& 990	& 110 \\
{[Ne~III]}                    &  8001.5	& 2.6  &   1.0683	&0.0006 & --	& --	& 1100	& 250 \\
\hline                                                                     
Si~III$^{*}$                  &   2474.1 & 1.5  &   1.0505      &0.0014   & -2590 & 200   & 1100  & 450 \\
Ly$\alpha$ (narrow abs.)     &	 2514.9	& 0.5  &   1.0695	&0.0004 & --	& --	& 500	& 70  \\
Ly$\alpha$ (narrow em.)      &	 2510.9	& 0.5  &   1.0662	&0.0004 & --	& --	& 1980	& 170 \\
Ly$\alpha$ (broad em.)       &	 2494.3	& 4.0  &   1.0525	&0.0041  & -2430	& 480	& 9300	& 500 \\
Ly$\alpha$ (broad abs.)      &	 2475.4	& 2.5  &   1.0368	&0.0020  & --	& --	& 7630	& 250 \\
N~V$^{*}$ (broad em.)              &  2540.0	& 10.0 &   1.0470	&0.0081  & -3100	& 1200	& 10000	& 800 \\
N~V (broad abs.)             &  2528.3	& 5.0  &   1.0376	&0.0040   & --	& --	& 3600	& 600 \\
\hline                                                                     
C~IV1550 (narrow abs.)        &  3208.7	& 0.3  &   1.0689	&0.0014  & --	& --	& 300	& 60  \\
C~IV1548 (narrow abs.)        &  3203.2	& 0.3  &   1.0689	&0.0005 & --	& --	& 300	& 60  \\
C~IV (broad em.)	      &  3180.9	& 4.4  &   1.0529	&0.0028  & -2300	& 400	& 10800	& 560 \\
C~IV (broad abs.)             &  3138.6	& 5.0  &   1.0256	&0.0032   & --	& --	& 6800	& 760 \\
C~IV? (narrow em)             &  3194.9	& 1.0  &   --	        &   --     & --	& --	& 1208	& 360 \\
\hline                                                                     
Al~III$^{*}$                   &  3828.9 & 2.8  &   1.0614      &0.0015   & -1030 & 220   & 3800 & 150 \\
C~III] (narrow em.)           &  3943.0	& 1.4  &   1.0660	&0.0005 & --	& --	& 1300	& 150 \\
C~III] (broad em.)            &  3926.6	& 3.5  &   1.0572	&0.0018  & -1640	& 270	& 7750	& 350 \\
\hline                                                                     
Mg II (broad em.)            &	 5748.3	& 6.9  &   1.0536	&0.0025  & -2160	& 360	& 13800	& 470 \\
Mg II (broad abs.)           &  5687.0	& 4.5  &   1.0317	&0.0016   & --	& --	& 5510	& 870 \\
Mg II? (narrow em.)          &  5775.6	& 1.6  &    --  	&  --        & -- 	& --	& 1490	& 330 \\
\hline\hline                                                               
\multicolumn{9}{c}{No broad absorption emission line best fit results} \\
\hline                                                                     
Ly$\alpha$ (broad em.)        &	 2501.4	& 2.1  &   1.0583	&0.0010  & -1470	& 250	& 8200	& 250 \\
C~IV (broad em.)	      &  3197.2	& 1.2  &   1.0634	&0.0008 & -739	& 110	& 10800	& 580 \\
Mg~II (broad em.)             &	 5775.0	& 1.7  &   1.0631	&0.0007  & -780	& 100	& 13500	& 440 \\
\end{tabular}
\tablefoot{The lines marked with an asterisk are not used to the average velocity offset. Question marks indicate 
emission lines of uncertain origin.}
\end{table*}

\subsubsection{Comparison with previous work}
\label{kuraszkiewicz}
The FOS spectrum was analyzed  in \cite{kuraszkiewicz02} as
part of  a large  catalog of  HST UV spectra  of QSOs.   Those authors
carried out a careful  fitting  of the UV  spectrum  of  3C~186 with  an
automated procedure that uses Gaussian components for all lines and no
broad absorption. They  derived a systemic redshift  of z=1.063, which
is inconsistent with  the value of z$_h$ we derived. The  redshift   derived  
from  the  Kuraszkiewicz  et
al. (2002)  model is lower  than the  redshift of the  isolated narrow
lines we  derived from the  SDSS spectrum  at a significance  level of
10$\sigma$.   However,  the optical  SDSS
spectrum  was   not  available for them to determine the accurate  value of
the systemic  redshift of the NLR  from the isolated narrow  [O~II] and
[Ne~III]  lines.   In Fig.~\ref{fig3}, we show the  wavelengths at which those lines would be
observed if  the redshift of the  target was z=1.063, as  estimated by
those authors  (the dot-dashed vertical  lines in  panel D). There  are no
detected emission lines at these  wavelengths, while the [O~II] and [Ne~III] 
lines are  clearly visible at the  wavelengths corresponding to
z$_h$.

\subsubsection{Is the QSO associated with an additional (undermassive) host?}
\label{nohost}
From the analysis of the spectroscopic data we are able to infer the virial BH mass of the QSO. 
We use the FWHM of the Mg~II line as an estimator  \citep{trakhtembrotnetzer12},
to limit any biases due to the possible contamination of winds that might affect high-ionization UV lines, and we obtain 
M$_{BH}$= $6\times 10^9$ solar masses.
We note that \citet{kuraszkiewicz02} estimated the black hole mass for this source using C~IV, and derived a similar mass
($3 \times 10^{9}$ M$_{\sun}$). The virial BH mass estimate is thus consistent with that based on the host galaxy magnitude (Sect.~\ref{bhmass}) 
within the uncertainties of the used relations. Furthermore, a rough lower limit on the black hole mass can be derived using the assumption that the BH accretes at
the Eddington limit. For L$_{\rm bol} \sim 10^{47}$, the lower limit on the BH mass is $8\times 10^8$ M$_\sun$.

The fact that 
such a value is consistent with the BH mass derived from the galaxy magnitude 
shows that the two objects perfectly match the expected properties
for a QSO and its host galaxy. However, for the purpose of providing a more
robust physical picture of the system,
it is important to firmly establish that a model that includes one PSF
and one galaxy  (hereinafter galaxy \#1) is the  best representation of
the HST  image and  thus the  host galaxy  we see  is the  only galaxy
associated with the  QSO. To this aim, we perform  a set of additional
tests  and simulations  using  {\it Galfit}. 

Firstly,  we  include a  second
Sérsic component (galaxy  \#2) in the model and fixed  its center to be
co-spatial with  the center of  the PSF.  When all the  parameters are
left  free  to vary  except  for  the centers  of  each  of the  three
components, the  fit does not  have a constrained solution  for galaxy
\#2.  The resulting  magnitude  of  galaxy \#2  is  only  a lower  limit
(m$_{\rm F140W} > 24.42$, i.e. $\sim$ 100 times fainter than the expected luminosity of
the host galaxy of  a $\sim 6\times 10^9$ M$_\sun$ BH), 
and the  resulting r$_{\rm eff}$ is $\sim$ 200\arcsec,
which corresponds  to $\sim$1.6Mpc at  the redshift of the  object. Clearly
this  solution  is unphysical.  Furthermore,  the  reduced $\chi^2$  worsens
significantly. Most notably, the properties of both the PSF and galaxy
\#1 are  unchanged with respect to  the best-fit model that  includes only two
components.  

Secondly,  we  add a  simulated  elliptical  galaxy
(S\'ersic  index  n=4,  ellipticity  e=0.5,  effective  radius  r$_{\rm eff}$=3\arcsec)
obtained using the package {\it artdata}  within IRAF to test whether {\it Galfit}
is able to correctly identify  the presence of an object significantly
fainter than  galaxy \#1. The  simulated galaxy is four  times (1.5mag)
fainter than  the detected host,  i.e. significantly fainter  than the
luminosity  expected  from  the  correlation  of  M$_{\rm BH}$  with  the  host
magnitude  \citep{marconihunt03},  even  taking into  account of  its
dispersion ($\sim$ 0.3 dex). When this additional component is superimposed onto the PSF,
{\it Galfit} is able to
correctly fit the image,  with a $\chi^2$ consistent
with that  of our  best-fit  model. Therefore, if  such a  galaxy were
present in the image,  {\it Galfit} would have been able to detect it. 

We also
simulate a  smaller (r$_{\rm eff}$=1\arcsec) spheroid  with the same magnitude  as in
the previous simulation, and we superimpose that to the QSO. {\it Galfit} is
again able to  fit such a component,  but with a larger  error on both
the effective  radius and the Sérsic  index. 
Therefore, even in the case of the lower BH mass estimate that was derived assuming accretion at the Eddington limit,
  the expected host galaxy would still lie within the range that we could detect with our method, based on the tests performed above.

We conclude that,  even if
the presence of an under-luminous object cannot be completely excluded,
the best model to fit the  image includes two components only: one PSF
and one  elliptical galaxy. Any  additional galaxy at the  location of
the PSF would be significantly under-luminous (possibly by a factor of
more  than 100,  according  to our  first test)  with  respect to  its
expected  luminosity, given  the BH  mass  of the  QSO. This  directly
implies that such a scenario is very unlikely.

\subsection{X-ray {\it Chandra}-ACIS data}
\label{chandra}
We analyzed archival {\it Chandra} X-ray observations to evaluate the possibility of a second AGN  
located  at the coordinates  corresponding to
the isophotal  center of  the host galaxy.  We register  the Chandra
0.5-7 keV  image to  the world  coordinate system  of the  HST WFC3-IR
image, using  the peaks of  the emission in  the two images  (i.e. the
position  of  the  unresolved  quasar). In registering the {\it Chandra} image to the framework of the HST image, we assume
that the bright point sources in each image are both associated with the QSO, and that they 
are co-spatial.

We  take  into  account  the
contributions from  the quasar (and  the associated PSF of  the ACIS-S
instrument), the cluster and the background.

\begin{figure}
\includegraphics[width=9cm]{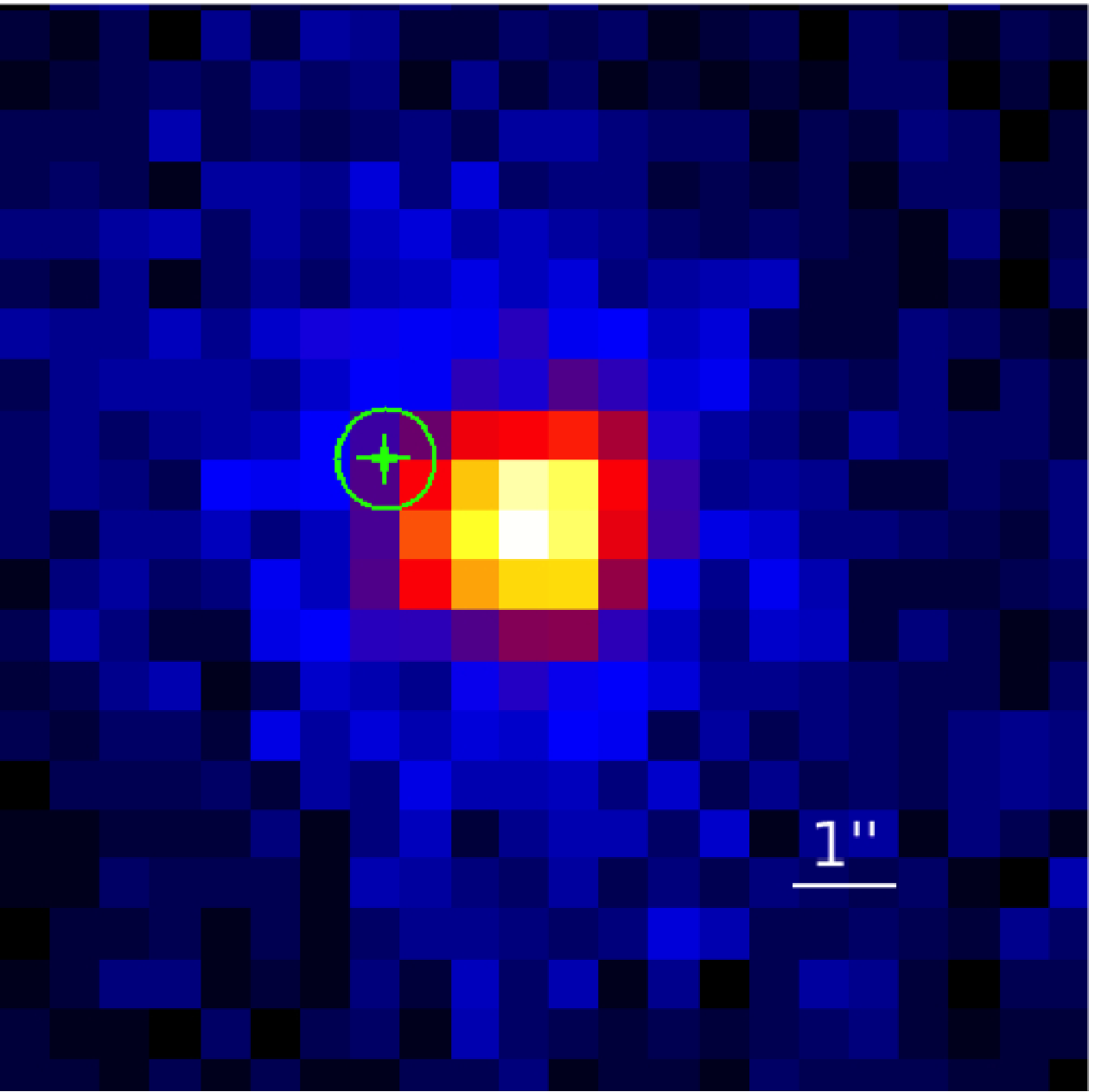}
\caption{Chandra X-ray  image of 3C~186.   The bright point  source at
  the center  of the field of  view is the quasar.   The region marked
  with the circle  corresponds to the location of  the galaxy nucleus,
  where the upper  limit for any additional AGN  source was estimated.
  In this figure North is up and East is left.}
\label{figchandra}
\end{figure}

We estimate a 3$\sigma$ upper limit F$_{\rm 2-10 keV}  < 2.2 \times 10^{-15}$ erg cm$^{-2}$ s$^{-1}$ for
a second AGN,  in a circular region with one  pixel radius centered at
the coordinates corresponding  to the host galaxy  center (Fig.~\ref{figchandra}).  This value
corresponds to  a 2-10 keV luminosity  L$_{\rm 2-10 keV} = 1.3\times 10^{43}$ erg  s$^{-1}$ at
the  redshift  of the  source.  This  is  $\sim$100  times lower  than  the
luminosity measured  for 3c186 (L$_{\rm 2-10 keV}=1.2\times 10^{45}$  erg s$^{-1}$). Assuming
that any  AGN at the  center of the host  is strongly obscured  in the
X-rays we  can correct its observed  X-ray flux for an  average factor
(Marinucci et al.  2012) of 70, leading to an  upper limit of L$^{\rm C-thick}_{2-10 keV}  
< 9.1 \times  10^{44}$ erg  s$^{-1}$. We also note  that the power  needed to
photo-ionize the  narrow emission lines  as estimated from  L$_{\rm [OIII]}$ \citep{hirst03} is
L$_{\rm 2-10 keV}=  7.3 \times 10^{45}$ erg  s$^{-1}$, using the relation  between these two
quantities (Heckman  et al. 2005).  This implies that if  the observed
QSO  and the  galaxy were  a chance  projection, the  presence at  the
center  of the  host of  an AGN  that is powerful enough  to photo-ionize  the
observed  narrow  lines  would  be  detected  at  $>3\sigma$  level,  even  if
Compton-thick absorption was present, under reasonable assumptions.

\section{Discussion}
\label{discussion}
We presented  evidence for  both a spatial  offset between  the nucleus of
3C~186 and its host galaxy, and  a velocity shift between the broad and
narrow emission lines in its spectrum.
In the following we discuss possible interpretations of our results.
We first consider scenarios that assume that the velocity offsets are caused by
peculiar properties of the central AGN,  i.e an extreme disk emitter or a peculiar wind. We then
discuss scenarios that can possibly account for both the velocity and spatial offsets displayed
by this source. We then outline the reasons why we believe that the interpretation
as a GW recoiling BH is favored when accounting for the overall properties of this object.

\subsection{An extreme disk emitter}
The velocity offsets observed in the spectra of 3C~186 can, in principle, be explained in
terms of extreme asymmetries of either the broad line region or the accretion disk.
One possibility is that 3C~186 represents an extreme case of an eccentric disk emitter \citep[e.g.][]{eracleous95}. 
Peculiar line profiles  and double-peaked broad low-ionization lines are 
found  in a fraction ($\sim 10\%$) of radio-loud  AGNs \citep[e.g.][]{eracleous03,strateva04,liu16}. 
Extreme cases of double-peaked lines in which the blue peak dominates 
may mimic the features observed in 3C~186.

However, there are a number of problems with such an interpretation:

i) eccentric disk models predict that the apparent velocity shifts observed
in low and high ionization lines (e.g. Mg~II and C~IV, respectively) should be different,
since the low-ionization lines are produced in the higher density accretion disk (thus originating the
double-peaked Balmer lines) while the higher ionization lines are thought to be produced in a low density
wind  \citep{murraychiang97,chenhalpern89,eracleous03,strateva04,braibant16}. 
In fact, Ly$\alpha$ and C~IV
are always single-peaked in objects with double-peaked Balmer and Mg~II lines \citep[e.g.][]{halpern96}.
On the other hand, the observations of 3C~186 presented here show that, for this object, 
all broad lines (of both low and high ionization) are shifted. Furthermore,
in the model in which broad absorption is used, all velocity offsets are consistent with the same value;

ii) objects with double-peaked and strongly
asymmetric  lines are known to show  significant variability  both in  the emission
line  profile  and  flux  \citep[e.g.][]{eracleous97,gezari07,liu16}, 
because of the intrinsic asymmetric structure of the accretion disk and the possible effects of winds.
The  spectra we  use in  our analysis  were taken  nine years
apart. The overlap between these two  spectra is small, since only the
C~III] line was observed at both epochs. The HST/FOS data are extremely
  noisy  in  the  C~III]  region,   and  do  not  add  any  significant
    information because a fit would be poorly constrained. However, we
    checked that the model parameters we  use to fit the C~III] line in
      the SDSS spectrum also provide a good representation of the same
      line in  the HST/FOC  data. We also note that the continuum emission in the region in which the
      spectra overlap is also consistent with being stable. While  we cannot  completely exclude
      that  variability is  present, it  is striking  that all  of the
      broad  lines  show  consistent  offsets,  considering  that  the two
      spectra were  taken nine  years apart.

\subsection{A peculiar wind}
The specific properties of the broad line offsets in 3C~186 could also be interpreted in the
  context of a wind scenario. The presence of winds is often used to explain significant blue-shifts  characterizing
  high-ionization lines in the spectra of quasars.
Using a large sample of SDSS quasars, \citet{shen16} show that high-ionization broad lines such as C~IV
are generally more blue-shifted than those of low ionization (such as Mg~II). 
In turn, low-ionization permitted  broad lines do not show large velocity offsets with respect to low ionization narrow lines
that are best used to derive the systemic redshift of the object (such as e.g. [O~II]).
Typical velocity offsets with respect to low ionization narrow lines are of the order of 
a few tens of km s$^{-1}$ for Mg~II, with an intrinsic spread of $\sim 200$ km s$^{-1}$.
Therefore, in most cases Mg~II can be considered as a good indication of the systemic redshift of the object.

Offsets displayed by C~IV with respect to Mg~II show strong 
luminosity dependence. However, such a relation is less strong and the velocity offsets are, on average, 
smaller for radio-loud QSOs \citep{richards11}.
\citet{shen16} show  that for bright QSOs, the average blue-shift of C~IV is $\sim 700$ km s$^{-1}$,
with a scatter of $\sim 100$ km s$^{-1}$.

While the properties observed  in the spectra of 3C~186 can be qualitatively explained in the framework of a disk-wind model,
the offsets we measure
are rather atypical, since we have both high (e.g. C~IV) and low (Mg~II) ionization lines that show similar blue-shifts. 
Velocity offsets as high as those that we measure in our source
are also extremely rare in the general QSO population.
This is particularly evident for low ionization lines such as Mg~II for which,
even using the spectral model that assumes no broad absorption, the derived velocity offset is $>3.5\sigma$ higher 
than those shown by the SDSS QSOs \citep{shen16}.

Based on these considerations, we conclude  that a scenario  that explains 
the data  with a peculiar  disk or disk+wind models cannot be completely ruled out, but is unlikely. 
Furthermore, we note that, while such a scenario might in principle account for 
the velocity offsets observed in some (but not all) of the emission lines,
it does not explain by itself the spatial offset observed in the image. 
To explain the properties of the spectrum of 3C~186 in the context of a disk/wind scenario, we would also
have to assume that the QSO is disconnected from the galaxy we see in the HST image, and it is in-falling towards that galaxy
at a velocity of the order of at least $\sim 1000$ km/s, to account for the velocity offset displayed
by the Mg~II line.

In the following we further discuss this type of  scenario, and we outline our favored interpretation for both the spatial and the velocity offsets
in the context of a single framework, without assuming any specific intrinsic peculiarity of the QSO.

\subsection{How do we explain both spatial and velocity offsets?}
\label{explain}
Four different scenarios may account for both intrinsic velocity  and  spatial offsets:
i) the  quasar and the  detected galaxy  are two
unrelated systems located  at different redshifts (with the quasar being a foreground object, 
as the broad lines are blue-shifted); in  this case, both
objects are AGNs (one produces the broad lines, the other one produces
the narrow lines only);  ii) same as above, but with the QSO as a background object that is moving towards the
detected galaxy;
iii) a  recoiling (slingshot) BH resulting from
the interaction  of a  double or  multiple BH system  in the  host, in
which at  least another BH  is active to produce  the narrow
lines (i.e. a dual or multiple AGN);  and iv)  a GW  recoiling black  hole.  

Regarding the first scenario, the detection  of UV absorption
lines in  the quasar spectrum  at a systemic velocity  consistent with
that of the  narrow emission lines directly implies  that the 
QSO cannot be  a foreground  object. 
The velocity offset derived using our model that includes broad absorption is significantly ($\sim 3\sigma$)
larger than the velocity dispersion of
the cluster  in which  the object resides  ($\sigma=780$ km/s),  as estimated
from the  properties of the  X-ray cluster emission  \citep{siemiginowska10}.  
The possibility that  the QSO is  a cluster object  in the
background that is in-falling towards the detected galaxy (second scenario) is thus
unlikely. If we assume that broad absorption  should not be used to fit the emission lines (but see Sect.~\ref{spectraresults} for a discussion on why we believe this would not be the best model), then the velocity offsets of all resonant lines (i.e. all but C~III]) are closer to the cluster velocity dispersion. Therefore, in this scenario, the possibility that the QSO is a background object, which is in-falling towards the galaxy we see in the HST image, cannot be rejected.
However, there are still two issues to be explained. Firstly,
the lack of a substantial host galaxy of the QSO, which should be significantly undermassive in order to be undetected,
as shown in  Sect.~\ref{nohost}. Secondly, and most importantly in such a scenario, the presence of two AGNs must be assumed, one that produces the broad lines, and one that produces the narrow line system at z$_h$. The presence of a Type~2 (hidden)
AGN at the center of the detected galaxy
could, in principle, explain the narrow line system at z$_h$. In this scenario, we would also expect to see a bright set of
narrow lines associated with the NRL of the QSO, at a redshift consistent with that of the broad emission lines. These lines are clearly absent from the spectrum (see Fig.~\ref{fig3}, panel D).

The interpretation of the data in terms of a dual or multiple AGN (third scenario) is also very
unlikely considering the power of the  ionizing source  needed to produce the observed
emission from the NLR.
This can be estimated from the luminosity  of the [O~III]5007 line
L$_{\rm [OIII]}=2.2 \times 10^{44}$ erg s$^{-1}$ \citep{hirst03}. 
We derive L$_{\rm bol}\sim 7.5  \times 10^{46}$ erg s$^{-1}$,
using appropriate  scaling relations  (Punsly \&  Zhang 2011).  This is
consistent with the luminosity  of the quasar L$_{\rm bol} \sim 10^{47}$ erg s$^{-1}$ 
as measured by \cite{siemiginowska10},  
and  directly implies  that the  QSO  is sufficient  to
photo-ionize  the observed  narrow  lines.  
Given these considerations, we infer that, not only an additional AGN is not required but, 
based on the analysis of {\it Chandra} data, we can also rule out  the
presence of  another powerful unobscured  or mildly obscured  
AGN located at the isophotal center of the host. In fact, the
upper limit  to the X-ray  emission of any  other accreting BH  at the
position corresponding to  the center of the host is  about two orders
of  magnitude lower  than the power  needed  to photo-ionize  the observed
narrow  emission lines.  The presence  of a  typical heavily  obscured
(Compton-thick) AGN is also very unlikely. In fact, the 3$\sigma$ upper limit
for the  X-ray intrinsic luminosity derived in Sect.~\ref{chandra} is  a factor of $\sim$8  lower than the
power  of the  Compton thick  AGN  needed to  photo-ionize the  narrow
lines.

Furthermore, the presence of additional SMBHs in the system is
unnecessary  to explain  the observations.  The BH  mass
derived using the host galaxy  magnitude converted to the (rest-frame)
infrared K-band as  an indicator \citep{marconihunt03} is M$_{\rm BH}=3.0\times 10^9$
M$_\sun$. The virial  mass estimate derived using the FWHM  of the Mg~II line
line \citep{trakhtembrotnetzer12} returns a similar value  ($6 \times 10^9$ M$_\sun$). Thus, 
the presence of another BH is not necessary, since the one associated with the QSO
has the mass we expect based on the properties of the host in which it resides.
In turn, this also implies that if the quasar resided in a host galaxy other than the one
detected in  the HST  image, the  host galaxy  associated with  such a
massive  BH  would be  detectable,  at  least  in the  model  residual
image, as shown in Sect.~\ref{nohost}. 

We note that a
dual AGN (or  AGN + inactive BH) scenario is also unlikely because of
the observed large velocity offset. Expected velocity offsets for dual AGNs are
of the  order of 10 to  100 km/s (e.g.  Wang \& Yuan 2012,  Comerford \&
Greene 2014). Furthermore, the Keplerian velocity  for a $\sim 10^9$ M$_\sun$ BH in
a binary  system, at  a distance  of $\sim$10kpc  from the  other component
would be of the order of a few tens of km s$^{-1}$, clearly inconsistent with
the observations.

Finally, the presence of a single host galaxy also disfavors a  
pre-merger black hole binary or, even
less likely, of a slingshot effect  due to a three-body interaction. In this scenario
we would expect to observe a galaxy merger still in progress. 

We conclude that the most likely explanation is the fourth scenario, which involves a GW recoiling black
hole. While we admittedly cannot completely exclude that an ad hoc combination of some (or all) of
the above discussed scenarios could conspire to give rise to the observed properties of our object, the GW recoiling BH scenario
naturally accounts for both the observed velocity and spatial offsets. Furthermore, it does not require any
ad hoc assumptions on the specific properties of host galaxy, BLR and NRL of this quasar.
In this scenario, 3C~186 would then be a normal QSO,
which simply happened to be ejected from its host galaxy by a well known mechanism that is expected, in some cases, as a result of a BH merger \citep[e.g.][]{peres62,loeb07,volonterimadau08}.

\subsection{A GW recoiling black hole}

Having explored a number of possible explanations to account for the observed properties of 3C~186, we favor
the GW recoiling BH scenario. The SMBH was likely ejected as a result of the gravitational radiation rocket  effect,
following a major  galaxy merger in which the SMBHs at the
center of  each merging galaxy also merged.  The accretion disk and  the broad
line region remained attached to the recoiling  BH. The narrow emission lines are
produced at larger distances from the BH with respect to the BLR, in the systemic frame of the host galaxy. 
This scenario explains the observed 
velocity offsets between the broad and narrow emission line systems. 
Furthermore, it also accounts for the
observation of the nuclear spatial offset with respect to the isophotal
center of the host galaxy.

\begin{figure}
\includegraphics[width=9cm]{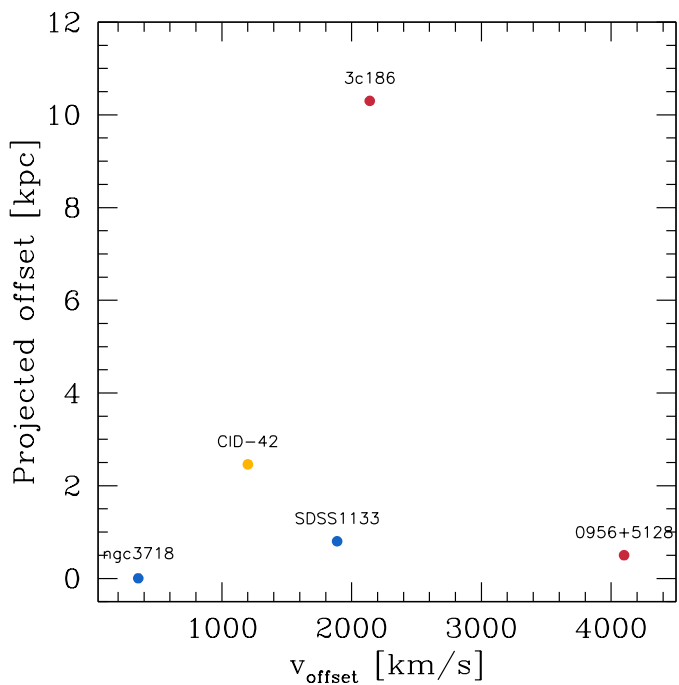}
\caption{Projected  spatial offset  plotted  against  broad to  narrow
  emission line  velocity offsets (absolute value)  in recoiling black
  hole candidates.  The color scale  reflects the bolometric  power of
  each object. Blue is used for  L$_{\rm bol} < 10^{43}$ erg s$^{-1}$,
  yellow for  $10^{43} <$ L$_{\rm bol}  < 10^{45}$ erg s$^{-1}$  and red
  for L$_{\rm  bol} > 10^{45}$  erg s$^{-1}$. For SDSS1133  the velocity
  offset  is   known  to  vary   between  ~40  km/s  and   ~1890  km/s
  \citep{koss14}.   In   this    figure   we   use the  measurement
  corresponding to the highest velocity. }
\label{plotgw}
\end{figure}

The measured velocity  offset is close to or slightly  higher than the
escape velocity expected for a  massive elliptical galaxy \citep[e.g.][]{merritt04}.  
High-velocity offsets (v$>$1000 km/s) are  expected to be
rare and they are more likely to be observed in combination with large
spatial offsets \citep{blecha16}. In Fig.~\ref{plotgw}, we show a comparison
between the velocity and nuclear offsets measured in some of the most interesting
GW recoiling BH candidates published so far.
We note that one of them (SDSS  0956+5128) displays even  larger velocity  shifts than 3C~186. 
However,  in this
case, the fact that the low-ionization broad permitted lines
(H$\beta$ and  Mg~II) show significantly different shapes  cannot  easily be
explained in terms of a recoiling  BH \citep{steinhardt12}. 3C~186 is one of the
highest velocity objects, but it is also the one object that shows the largest
spatial offset.

\subsubsection{Relevant timescales and effects on the observed radio and optical morphologies}
The first important timescale we can derive from the observations is the time since  
the GW kick was received as the two BHs merged.
Given the measured velocity of $\sim 2100$ km s$^{-1}$ and the 1.3'' spatial offset, we derive that the
time since  the BH  merger event  is $\sim5$Myr,  assuming both a constant velocity and that the angle
between the direction  of motion and the line-of-sight  is $\sim$45deg. 

An important property of 3C~186 is 
that it possesses  powerful relativistic jets.
The presence of radio jets allows an estimate of the age of the SMBH activity, based on
the synchrotron radiative cooling timescale.
\cite{murgia99} estimated a radiative age of $\sim 10^5$ years for 3C~186. This implies that the 
radio AGN turned on at a  later time with respect to the time of the GW kick.
We note  that since the radio source is very
young we do not expect to see any significant bending in the radio jet
as a result of the BH motion. Assuming a projected velocity of the order of 
that measured from the spectra ($\sim 2100$ km $^{-1}$),
any displacement of the hot-spots with respect to radio core would be less than 0.1”.
This is consistent with the observed radio morphology, in which the jet appears 
roughly straight \citep{spencer91}. 
However, the hot-spots are slightly displaced with respect to the jet direction, displaying 
an S-shaped morphology. This type of radio morphology is usually interpreted as being due to a jet emitted 
by  a precessing BH \citep{eckers78},
which is to be expected as a result of a BH merger with misaligned spins and/or uneven BH masses.

We can also estimate the lifetime of
an accretion disk attached to a BH kicked at a velocity 
$\sim 2100$ km s$^{-1}$. Using the formula
in \cite{loeb07}, and assuming a radiative efficiency of $\epsilon=0.1$ and 
the luminosity and BH mass estimated for 3C~186, we derive t$_{\rm disk}\sim 10^8$ yr.
This is a significantly longer timescale than the time since the GW kick occurred.
This implies that the accretion disk can survive until the BH reaches very large 
distances from the center of the host,
thanks to the fact that its lifetime strongly depends on the BH mass.

The host galaxy shows the  presence of low surface brightness features
in  its  outer  regions,  possibly  shells or  tidal  tails  that  are
typical  of  major galaxy merger remnants, 
i.e. those in which the two merging galaxies have
masses that are equal to within a factor of 3 (Fig.~\ref{fig1},  bottom-right
panel). From a qualitative  comparison with simulations \citep[e.g.][]{springel05,lotz08}, 
we  estimate that the  galaxy merger
event happened  on timescales of  about 1Gyr  or more, since  only one
distinct   galaxy    with   a   relatively   smooth    morphology   is
visible.  Furthermore, the  S\'ersic  index resulting  from  the fit  is
consistent with  that of  relaxed elliptical galaxies.  

A quantitative
comparison is extremely difficult with the current data because of the
complexity of the problem,  as well as the low S/N of  the image in the
outer regions of the host galaxy, which prevents us from disentangling the
faint structures in the possible  tidal tails. Simulations (e.g. Lotz et al. 2008) show
that the expected morphologies at  different times since the beginning
of  the  merger  are  strongly dependent  on  the  initial  parameters
(i.e.  mass, gas  content, galaxy  morphology). However,  it is  clear
(e.g. Figs. 1 and 2 in Lotz et al. 2008) that for t$<$1-2Gyr the central
regions of the  merged galaxy are still  significantly disturbed. This
is not what we  observe in the host of 3C~186, where  the galaxy can be
accurately modeled  with a  smooth Sérsic component. 
Thus, the galaxy merger must have occurred more than 1-2 Gyr
ago.

It  is possible  that when  the  target is  observed  with higher  resolution
instruments we may be  able to see more  details in the
innermost kpcs. But  since the spatial resolution  offered by HST/WFC3
is $\sim 0.8$ kpc at the redshift of  the source, we believe that significant
disturbances would be  visible in our data (see, for example, the morphologies
of some  of the merging  3C radio  galaxies presented in  Chiaberge et
al. 2015).

Finally, we note that in a galaxy merger in which both galaxies possess an SMBH, 
the timescale for two SMBHs to sink into the center of the merger remnant and form a bound binary is likely 
at least one order of
magnitude  shorter than the timescale of 1-2 Gyr (or more)
that we roughly estimate for the galaxy merger \citep[e.g.][]{begelman80,khan12}. 
This implies that if we assumed that the two BHs in the 3C~186 merging system have not merged yet, and that what we are observing is 
a SMBH binary system, the observed large velocity offsets would be inconsistent with the small 
velocities expected for a BH binary (see Sect.~\ref{explain}).

\section{Conclusions}
\label{conclusions}

Irrespective of the specific interpretation of the results, 
3C~186 is  an extremely  interesting and  unique object. 
We measure a spatial offset of $1.3''$ between the QSO point source and 
the isophotal center of the host galaxy. This
corresponds to a projected distance of $\sim 11$kpc at the redshift of the source.
The broad emission lines show significant velocity offsets (v$\sim 2100$ km s$^{-1}$) with respect to the
systemic redshift of the host galaxy as derived from the narrow emission and absorption line system.
We showed that
the most plausible explanation for both the nuclear spatial offset seen in
the HST/WFC3-IR image and in the spectra is in terms of a gravitational wave
recoiling black hole scenario, although the explanation in terms of a peculiar
  background QSO associated with an undermassive host galaxy and/or characterized by peculiar
  winds cannot be completely excluded.
Based on the morphology of the host, we estimate that 
a major merger between two galaxies both containing an SMBH 
occurred roughly $1-2$ Gyr ago. When the BH-BH merger occurred, probably $\sim 5 \times 10^{6}$ years ago based 
on both the observed velocity and spatial offsets,
the anisotropic emission of gravitational waves generated a kick that ejected the merged SMBH from the
central region of the merged host galaxy. The AGN accretion disk remained attached to the
BH, thus causing the observed velocity offsets of the broad emission lines with respect to the NLR.
Spectral aging arguments show that the radio-loud AGN turned 
on more recently, $\sim 10^{5}$ years ago \citep{murgia99}. 
Theoretical considerations \citep{loeb07} have been made 
that indicate that the accretion disk can survive in such 
a condition for a timescale of as long as $\sim 10^{8}$ years.

3C~186 is a
perfect laboratory to study all  of the effects associated with galaxy
and BH  mergers, the timescales  involved in these processes,  and the
production of gravitational waves. 
The fact that this object is radio-loud is extremely interesting. On the one hand, some of  the best models  
to explain the production of relativistic jets require the presence  of a
rapidly spinning BH \citep[e.g.][]{blandfordznajek77,mckinney12,ggnature14}. 
On the other hand, there  is growing evidence that RLAGNs are
closely  linked to  major galaxy  and BH  mergers 
\citep{wilsoncolbert95,ivison12,ramosalmeida13,papmergers}. Interestingly, one possible way to spin-up
the   black  hole   is  via   a  BH-BH   merger  with   specific  spin
configurations \citep[e.g.][]{schnittman13,hemberger13}. Therefore, it is not surprising
that we are able to observe a GW recoiling SMBH associated with a radio-loud AGN.

A number  of future
studies of 3C~186 should be performed 
to further investigate the properties of this intriguing source and test the
proposed GW recoiling black hole scenario. 
Deeper HST images will allow us to
both completely rule out the  presence of an under-massive host galaxy
around  the QSO,  and to  better study  the properties  of the galaxy merger
remnant. This should include color information, to determine the age of the stellar populations 
at different locations in the host galaxy, and set constraints on the galaxy merger timescales.
Spectroscopy at UV and optical wavelengths will enable the monitoring of
the spectral features we observed, which is crucial to determine whether these are transient or permanent phenomena.
High-resolution spectroscopy with high S/N 
will also enable a more accurate measurement of the line offsets.
Observing the spectral region of the H$\beta$ emission line would provide a cleaner picture of the low-ionization
BLR, since such a line is significantly less contaminated by other spectral features with respect to the Mg~II UV line and the other lines presented here.
If the GW recoiling BH
scenario holds, we expect the broad H$\beta$ line to show a velocity offset consistent with those measured
in the lines presented in this paper (i.e. $\sim 2000$ km/s).
IFU data taken with an 8m-class telescope and adaptive optics 
will enable us to identify the spatial location of the NLR, and 
to observe the expected decoupling of the BLR and NLR.

Furthermore, ALMA observations will enable  the study of the kinematics of
the  molecular gas  in the  vicinity of  the recoiling  BH and  at the
center of the host galaxy. Future X-ray observations with space-based 
X-ray high-resolution spectrographs, such as the one that will be on 
board the planned ESA X-ray telescope Athena, will allow direct measurements 
of (or put stringent limits on) the BH spin in this source.    
Numerical modeling will also be extremely important  to better
understand important  issues, from the triggering  mechanisms of radio
loud AGNs  to their connection with  merging BHs.   

Finally, if the GW recoiling BH scenario is correct,
these results are clearly also relevant for gravitational wave studies.
GWs from  $\sim$30 M$_\sun$ merging BHs  were recently detected by  LIGO (Abbott et al.  2016) and
pulsar timing  experiments (EPTA,  PPTA, and SKA  in the  future) will
in the future be able to detect  GWs of the  frequency expected  from mergers of  BHs with
masses similar to that of 3C~186 (Moore et al. 2015, Babak et al. 2016,
Madison et al. 2016).

\begin{acknowledgements}
The authors thank Julian Krolik, Tim Heckman, Marta Volonteri and Ski Antonucci for providing insightful comments.  J.P.K. and B.H. acknowledge support from HST-GO-13023.005-A. We thank the anonymous referee for their comments that helped to improve the paper. This work is based on observations made with the NASA/ESA HST, obtained from the Data Archive at the Space Telescope Science Institute, which is operated by the Association of Universities for Research in Astronomy, Inc., under NASA contract NAS 5-26555. This research has made use of the NASA/IPAC Extragalactic Database (NED) which is operated by the Jet Propulsion Laboratory, California Institute of Technology, under contract with the National Aeronautics and Space Administration.
 \end{acknowledgements}

%
%

\begin{appendix}
  \section{Palomar TripleSpec data: preliminary results}
\label{palomar}

  \subsection{The data}
  We obtained a spectrum of 3C~186 with the Palomar 200'' Telescope and TripleSpec in the J, H and K near-IR bands. The spectrograph sensitivity curve extends from $\sim$
  1.0 to 2.4 $\mu$m, and the spectral resolution varies between 2500 and 2700, depending on the wavelength. The data were
  taken on December 10-12, 2016. Here we present data from the first part of the observation, and we focus on the J- and K-band spectra, which include
  information relevant to this paper. The full dataset will be presented in a forthcoming paper.
  The exposure time was 245 minutes,using an A-B-B-A sequence. Calibration data were also
  taken during the same nights, including flats, darks and calibration stars for telluric subtraction and flux calibration. The spectrum was reduced
  and calibrated using SpexTool \citep{spextool}. Telluric absorption bands cover spectral regions that, at the redshift of the source,
  include both H$\alpha$ and H$\beta$. Therefore, the major broad emission lines that fall in the near-IR, as accessible from ground-based
  telescopes, are He~I 10830 and Paschen $\gamma$ \citep[see e.g.][]{riffel06}.
  Paschen $\delta$ is also detected, but it is significantly fainter and falls in a noisier region of the spectrum. These lines all fall in order 3
  of the spectrum, which corresponds to a spectral region between $\sim$ 1.9 and 2.4 $\mu$m. Order 6 of the spectrum includes the [O~III]5007,4959 doublet.

\begin{figure}
\includegraphics[width=8cm]{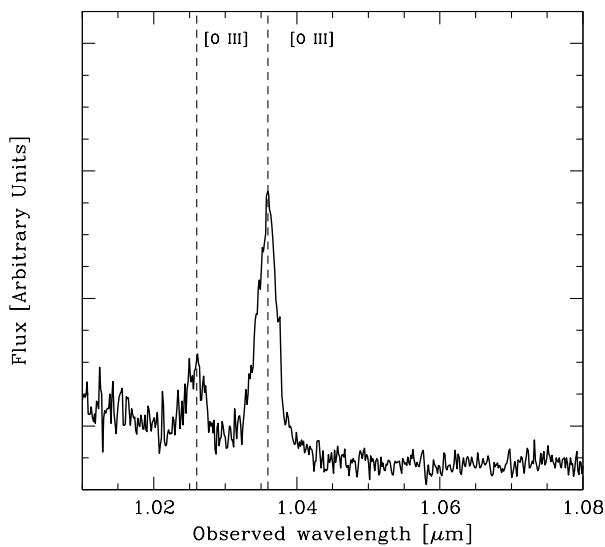}
\caption{Palomar TripleSpec spectrum (order 6). Wavelengths are in the observer’s frame. The [O~III]5007,4959 doublet is shown. The vertical dashed lines indicate
  the wavelengths of these two lines corresponding to the systemic redshift z$_h = 1.0685$.}
\label{palospec}
\end{figure}

\begin{figure}
\includegraphics[width=8cm]{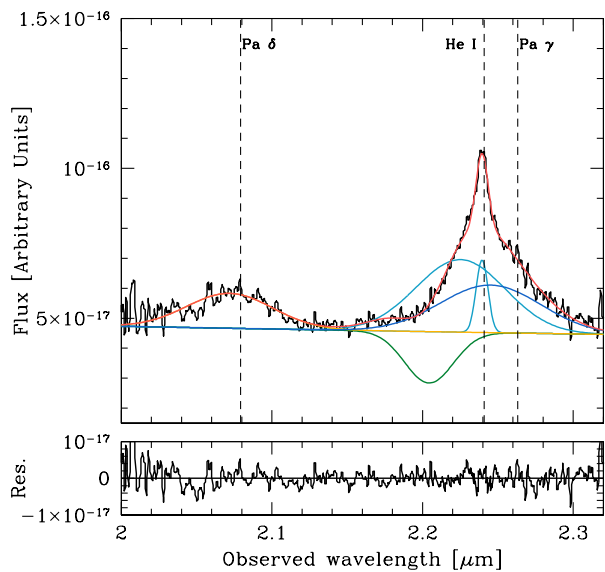}
\caption{Palomar TripleSpec spectrum (order 3). Wavelengths are in the observer’s frame. Relevant lines are labeled at the top of the panel.
  The vertical dashed lines indicate
  the wavelengths of the lines corresponding to the systemic redshift z$_h = 1.0685$.
  The best fit is the red line. Each component of the model is shown separately, added to the continuum power law, for clarity. The emission components are shown in cyan and blue for He~I and Pa $\gamma$, respectively. The absorption component is plotted in green. Model residuals are shown in the bottom panel.}
\label{palospec2}
\end{figure}

  \subsection{Results}
  The relevant segments of the Palomar TripleSpec spectrum are shown in Figs.~\ref{palospec} and \ref{palospec2}.
  In Fig.~\ref{palospec} we show the spectral region around the [O~III]5007,4959 doublet.
  The dashed vertical lines show the wavelengths of these two emission lines corresponding to the systemic redshift z$_h=1.0685$. It is clear that both lines
  are centered at z$_h$, in agreement with the [O~II] and [Ne~III] lines detected in the SDSS spectrum (see Sect.~\ref{spectraresults}). Fitting this
  region is rather complex  because of the presence of the red side of the broad H$\beta$ line, visible as a rising component blue-ward of
  the [O~III]4959 line. Unfortunately, atmospheric absorption
  combined with the reduced sensitivity of the spectrograph at  $\lambda < 1 \mu$m do not allow  accurate modeling of the H$\beta$ line, thus both the peak
  wavelength and width are unconstrained. However, the central wavelengths of the [OIII] emission lines can be accurately measured.
  In Fig. \ref{palospec2} we show the spectral region of the He~I line. The line is blended with Paschen $\gamma$. The Paschen $\delta$ line is also detected, but
  it is significantly fainter. The shape of the strongest line (He~I) is similar to that of the UV broad resonant lines (i.e. the blue side of
  the line shows a   concave profile). 

  We model the lines using the same technique as described in Sect.~\ref{spectralmodeling}. In Table~\ref{tabpalomar} we report the results of
  the best fit. In model A, we adopt a set of narrow and broad emission lines with one broad absorption component for He~I, in analogy
  with the UV broad resonant lines and to better reproduce the concave line profile. He~I and Paschen $\gamma$  are
  offset by $\sim 2200$ km/s  with respect to the  systemic redshift. The Paschen $\delta$ line shows a smaller offset (i.e. 1050 km/s).
  However, the signal-to-noise for this line is significantly lower because of the presence of the strong carbon dioxide
  absorption features at $\sim$2.06 $\mu$m. Furthermore, for such a line we do not include a broad absorption component.
  In model B we fit the He~I complex without including a broad absorption component. Significant offsets are still present ($\sim 1500$ km/s). However,
  the quality of the fit is significantly lower and the errors are larger.
  We compared the $\chi^2$ derived for each model and we obtain that the model with broad absorption
  better represents the data with a high level of statistical significance (P $<<$ 0.001). We note that He~I also shows a narrow emission component that
  is offset by  $\sim 200$ km/s, again in agreement with the narrow components observed in the UV resonant lines (see Tab.~\ref{table2}).

  We conclude that these preliminary results support the presence of significant velocity offsets, in quantitative agreement with those
  derived for the UV emission lines in Sect.~\ref{spectraresults}.

\begin{table*}
\caption{Emission lines best fit model parameters, Palomar TripleSpec data}             
\label{tabpalomar}      
\centering          
\begin{tabular}{l c c c c c c c c }     
\multicolumn{9}{c}{Model A: With broad absorption component} \\
\hline\hline       
Line component	      & Observed wavelength & Err.            &  Redshift &   Err. &  Velocity offset  &   Err. & FWHM &	Err \\
   ~                  & $\lambda$ [\AA])    &  ~ & $z$ & ~ & [km s$^{-1}$] & ~ & [km s$^{-1}$] \\
\hline  
\hline
He~I (broad em.)             & 22252  & 8.3  &   1.0540       &0.0004 & -2100	& 110	& 9400	& 150 \\
He~I (broad abs.)            & 22042  & 4.5  &   1.0347       &0.0002 & --	& --	& 5200	& 60  \\
He~I (narrow em.)            & 22395  & 1.0  &   1.0672       &0.0001 & -190	& 10	& 1200	& 50  \\
Paschen $\gamma$             & 22450  & 9.8  &   1.0520       &0.0004 & -2400	& 130	& 10500	& 100 \\
Paschen $\delta$             & 20720  & 5.1  &   1.0612       &0.0003 & -1100   & 200   & 9700  & 250 \\
\hline\hline                                                               
\multicolumn{9}{c}{Model B: No broad absorption component} \\
He~I (broad em.)             &  22309	& 22.0  &   1.0593       &0.0010 & -1320 & 300	& 11200	& 300 \\
He~I (narrow em.)            &  22382	& 1.5   &   1.0686       &0.0001 & -360	& 10	& 2400	& 50 \\
Paschen $\gamma$             &  22502	& 40.0  &   1.0566	&0.0020  & -1700& 500	& 9100	& 100 \\
Paschen $\delta$             &  20720   & 6.0   &   1.0613      &0.0003  & -1050 & 90   & 11100 & 200 \\
\hline                                                                     
\end{tabular}
\end{table*}

\end{appendix}

\end{document}